\documentclass[12pt]{article}
\usepackage{amsmath,amsfonts,amssymb}

\begin{document}
\thispagestyle{empty}

\def\theequation{\arabic{section}.\arabic{equation}}
\def\a{\alpha}
\def\b{\beta}
\def\g{\gamma}
\def\d{\delta}
\def\dd{\rm d}
\def\e{\epsilon}
\def\ve{\varepsilon}
\def\z{\zeta}
\def\B{\mbox{\bf B}}

\newcommand{\h}{\hspace{0.5cm}}

\begin{titlepage}
\vspace*{1.cm}
\renewcommand{\thefootnote}{\fnsymbol{footnote}}
\begin{center}
{\Large \bf Close to the Giant Magnons}
\end{center}
\vskip 1.2cm
\centerline{\bf Plamen Bozhilov}
\vskip 0.6cm
\centerline{\sl Institute for Nuclear Research and Nuclear Energy}
\centerline{\sl Bulgarian Academy of Sciences} \centerline{\sl  1784
Sofia, Bulgaria}

\centerline{\tt plbozhilov@gmail.com}

\vskip 20mm

\baselineskip 18pt

\begin{center}
{\bf Abstract}
\end{center}
\h We consider the most general string configurations on the
$R_t\times S^3$ subspace of $AdS_5\times S^5$, described by the
Neumann-Rosochatius integrable system. Under some restrictions on
the parameters of the solution and in an appropriate limit, they
correspond to small deviation from the known finite-size giant
magnon solutions with one and two angular momenta. Analyzing the
finite-size effect on the dispersion relation, we find that the
leading correction is modified in a way similar to the
$\gamma$-deformed case $R_t\times S^3_{\gamma}$. The subleading
correction for a string with one angular momentum is also found. It
depends on the same parameter, which describes the generalization of
the leading correction.

\end{titlepage}
\newpage

\def\nn{\nonumber}
\def\tr{{\rm tr}\,}
\def\p{\partial}
\newcommand{\bea}{\begin{eqnarray}}
\newcommand{\eea}{\end{eqnarray}}
\newcommand{\bde}{{\bf e}}
\renewcommand{\thefootnote}{\fnsymbol{footnote}}
\newcommand{\be}{\begin{equation}}
\newcommand{\ee}{\end{equation}}

\vskip 0cm

\renewcommand{\thefootnote}{\arabic{footnote}}
\setcounter{footnote}{0}


\setcounter{equation}{0}
\section{Introduction}
The AdS/CFT duality \cite{AdS/CFT} has become a major subject of
investigations in contemporary high energy physics. The most
developed part of this correspondence is between type IIB string
theory on $AdS_5\times S^5$ and $\mathcal{N}=4$ super Yang-Mills
(SYM) in four dimensions. Recent developments in this direction are
mainly based on the integrable structures discovered in both
theories.

Integrability on the SYM side appears in the calculations of
conformal dimensions, which are related to the string energies
according to the AdS/CFT correspondence. The remarkable observation
by Minahan and Zarembo \cite{MZ} is that the conformal dimension of
an operator composed of the scalar fields in ${\cal N}=4$ SYM, in
the planar limit, can be computed by diagonalizing the Hamiltonian
of one-dimensional integrable spin chain. This result has been
further extended to the full $PSU(2,2|4)$ sector \cite{Beis} and the
Bethe ansatz equations which are supposed to hold for all loops have
been conjectured \cite{BDS,BeiStai}.

The type IIB string theory on $AdS_5\times S^5$ is described by a
nonlinear sigma model with $PSU(2,2|4)$ symmetry \cite{MetTse}. This
sigma model has been shown to have an infinite number of local and
nonlocal conserved currents \cite{Bena}, and some of the conserved
charges such as energy and angular momenta have been computed
explicitly from the classical integrability. These results based on
the classical integrability provide valuable information on the
AdS/CFT duality in the domain of large t'Hooft coupling constant.

Various classical string solutions played an important role in
testing and understanding the AdS/CFT duality in the case under
consideration. The classical giant magnon \cite{HM06} discovered in
$R_t\times S^2$ gave a strong support for the conjectured all-loop
$SU(2)$ spin chain and made it possible to get a deep insight in the
AdS/CFT duality. In addition, this solution is related to the
integrable sine-Gordon model. It was extended to the dyonic giant
magnon, which corresponds to a string moving on $R_t \times S^3$ and
related to the complex sine-Gordon model \cite{Dorey,CDO06}. Further
extensions to $AdS_5 \times S^5$ have been also worked out
\cite{KRT06,SprVol,Ryangi,DimRas}.

The reduction of the classical string dynamics to the
Neumann-Rosochatius (NR) integrable system \cite{AFRT,ART,KRT06},
has proved to be an useful tool for describing a large class of
string solutions on $AdS_5\times S^5$ and other backgrounds,
including giant magnons and spiky strings. It can be used also for
studding the finite-size effects, related to the wrapping
interactions in the dual field theory \cite{Janikii}.

The finite-size effect for the giant magnon has been computed from
the $S$-matrix in \cite{Janikii} and has been shown to be consistent
with the classical string result to the leading order. The
finite-size effect for the giant magnon has been first found by
solving the string sigma model in a uniform and conformal gauges
\cite{AFZ06} and, subsequently, many related results, such as gauge
independence \cite{AFGS}, multi giant magnon states \cite{MinSax}
and quantization of finite-size giant magnon \cite{RamSem}, have
been derived. This result has been also related to explicit
solutions of the sine-Gordon equation in a finite-size space
\cite{KMcL08}. The dispersion relation for the finite-size dyonic
giant magnon was obtained in \cite{HS08}.


In this article, we will consider string configurations on the
$R_t\times S^3$ background, which depend on one more parameter,
compared to infinite or finite-size giant magnons with one or two
angular momenta. Our aim is to find the corresponding
generalizations of the finite-size corrections to the dispersion
relations. To this end, in section 2 we introduce the classical
string action on $R_t\times S^3$ and the corresponding NR system. In
section 3 we compute the conserved quantities and angular
differences of interest. In section 4 we obtain the dispersion
relation for the finite-size dyonic string up to the leading order.
Section 5 is devoted to the derivation of the subleading correction
to the energy-charge relation for the finite-size string with one
nonzero angular momentum. We conclude the paper with some remarks in
section 6. Appendix A contains information about the elliptic
integrals appearing in the calculations, the $\epsilon$-expansions
used and the solutions for the parameters.

\setcounter{equation}{0}
\section{Strings on $R_t\times S^3$ and the NR Integrable System}
Here we will briefly remaind how the string dynamics on $R_t\times
S^3$ can be reduced to the one of the NR integrable system
\cite{AFRT,ART,KRT06}. We begin with the Polyakov string action \bea
&&S^P= -\frac{T}{2}\int d^2\xi\sqrt{-\gamma}\gamma^{ab}G_{ab},\h
G_{ab} = g_{MN}\p_a X^M\p_bX^N,\\ \nn &&\p_a=\p/\p\xi^a,\h a,b =
(0,1), \h(\xi^0,\xi^1)=(\tau,\sigma),\h M,N = (0,1,\ldots,9),\eea
and choose {\it conformal gauge} $\gamma^{ab}=\eta^{ab}=diag(-1,1)$,
in which the Lagrangian and the Virasoro constraints take the form
\bea\label{l}
&&\mathcal{L}_s=\frac{T}{2}\left(G_{00}-G_{11}\right) \\
\label{00} && G_{00}+G_{11}=0,\qquad G_{01}=0.\eea

We embed the string in $R_t\times S^3$ subspace of $AdS_5\times S^5$
as follows \bea\nn Z_0=Re^{it(\tau,\sigma)},\h
W_j=Rr_j(\tau,\sigma)e^{i\phi_j(\tau,\sigma)},\h
\sum_{j=1}^{2}W_j\bar{W}_j=R^2,\eea where $R$ is the common radius
of $AdS_5$ and $S^5$, and $t$ is the $AdS$ time. For this embedding,
the metric induced on the string worldsheet is given by \bea\nn
G_{ab}=-\p_{(a}Z_0\p_{b)}\bar{Z}_0
+\sum_{j=1}^{2}\p_{(a}W_j\p_{b)}\bar{W}_j=R^2\left[-\p_at\p_bt +
\sum_{j=1}^{2}\left(\p_ar_j\p_br_j +
r_j^2\p_a\phi_j\p_b\phi_j\right)\right].\eea The corresponding
string Lagrangian becomes \bea\nn \mathcal{L}=\mathcal{L}_s +
\Lambda_s\left(\sum_{j=1}^{2}r_j^2-1\right),\eea where $\Lambda_s$
is a Lagrange multiplier. In the case at hand, the background metric
does not depend on $t$ and $\phi_j$. Therefore, the conserved
quantities are the string energy $E_s$ and two angular momenta
$J_j$, given by \bea\label{gcqs} E_s=-\int
d\sigma\frac{\p\mathcal{L}_s}{\p(\p_0 t)},\h J_j=\int
d\sigma\frac{\p\mathcal{L}_s}{\p(\p_0\phi_j)}.\eea

In order to reduce the string dynamics to the NR integrable system,
we use the ansatz \cite{KRT06} \bea\label{NRA}
&&t(\tau,\sigma)=\kappa\tau,\h r_j(\tau,\sigma)=r_j(\xi),\h
\phi_j(\tau,\sigma)=\omega_j\tau+f_j(\xi),\\ \nn
&&\xi=\alpha\sigma+\beta\tau,\h \kappa, \omega_j, \alpha,
\beta=constants.\eea It can be shown that after integrating once the
equations of motion for $f_j$, which gives (prime is used for
$d/d\xi$) \bea\label{fafi} f'_j=\frac{1}{\alpha^2-\beta^2}
\left(\frac{C_j}{r_j^2}+\beta\omega_j\right),\h C_j=constants, \eea
one ends up with the following effective Lagrangian for the
coordinates $r_j$ \bea\label{LNR} L_{NR}=(\alpha^2-\beta^2)
\sum_{j=1}^{2}\left[r_j'^2-\frac{1}{(\alpha^2-\beta^2)^2}
\left(\frac{C_j^2}{r_j^2} + \alpha^2\omega_j^2r_j^2\right)\right]
+\Lambda_s\left(\sum_{j=1}^{2}r_j^2-1\right).\eea This is the
Lagrangian for the NR integrable system \cite{KRT06}.

The  Virasoro constraints (\ref{00}) give the conserved Hamiltonian
$H_{NR}$ and a relation between the embedding parameters and the
integration constants $C_j$: \bea\label{HNR}
&&H_{NR}=(\alpha^2-\beta^2)
\sum_{j=1}^{2}\left[r_j'^2+\frac{1}{(\alpha^2-\beta^2)^2}
\left(\frac{C_j^2}{r_j^2} + \alpha^2\omega_j^2r_j^2\right)\right]
=\frac{\alpha^2+\beta^2}{\alpha^2-\beta^2}\kappa^2,
\\ \label{01R} &&\sum_{j=1}^{2}\omega_j C_j + \beta\kappa^2=0.\eea
On the ansatz (\ref{NRA}), $E_s$ and $J_j$ defined in (\ref{gcqs})
take the form \bea\label{cqs} E_s=
\frac{\sqrt{\lambda}}{2\pi}\frac{\kappa}{\alpha}\int d\xi,\h J_j=
\frac{\sqrt{\lambda}}{2\pi}\frac{1}{\alpha^2-\beta^2}\int d\xi
\left(\frac{\beta}{\alpha}C_j+\alpha\omega_j r_j^2\right),\eea where
we have used that the string tension and the 't Hooft coupling
constant $\lambda$ in the dual $\mathcal{N}=4$ SYM are related by
\bea\nn TR^2=\frac{\sqrt{\lambda}}{2\pi}.\eea

\setcounter{equation}{0}
\section{Conserved Quantities and Angular Differences}

If we introduce the variable \bea\nn \chi=1-r_1^2=r_2^2,\eea and use
(\ref{01R}), the first integral (\ref{HNR}) can be rewritten as
\bea\nn &&\chi'^{2}= \frac{4\omega_1^2(1-u^2)}{\alpha^2(1-v^2)^2}
\left\{-\chi^3+\frac{2-(1+v^2)W-u^2}{1-u^2}\chi^2\right.\\
\nn &&-\left. \frac{1-(1+v^2)W+[(v W-u K)^2-K^2]}{1-u^2}\chi-
\frac{K^2}{1-u^2}\right\}
\\ \label{eqchi} &&=\frac{4\omega_1^2(1-u^2)}{\alpha^2(1-v^2)^2}
(\chi_{p}-\chi)(\chi-\chi_{m})(\chi-\chi_{n}) ,\eea where \bea\nn
v=-\frac{\beta}{\alpha},\h u=\frac{\omega_2}{\omega_1},\h
W=\left(\frac{\kappa}{\omega_1}\right)^2,\h
K=\frac{C_2}{\alpha\omega_1}.\eea We are interested in the case when
\bea\nn 0<\chi_{m}<\chi< \chi_{p}<1,\h \chi_{n}<0.\eea The three
equations following from (\ref{eqchi}) are \bea\nn
&&\chi_p+\chi_m+\chi_n=\frac{2-(1+v^2)W-u^2}{1-u^2},\\ \label{3eqs}
&&\chi_p \chi_m+\chi_p \chi_n+\chi_m \chi_n=\frac{1-(1+v^2)W+(v W-u
K)^2-K^2}{1-u^2},\\ \nn && \chi_p \chi_m \chi_n=-
\frac{K^2}{1-u^2}.\eea We will use them essentially to find the
dispersion relations for the string configurations under
consideration.

Correspondingly, the conserved quantities (\ref{cqs}) transform to
\bea\nn &&\mathcal{E}\equiv
\frac{E_s}{\frac{\sqrt{\lambda}}{2\pi}}=\frac{\kappa}{\alpha}\int_{-r}^{r}d\xi
=\frac{(1-v^2)\sqrt{W}}{\sqrt{1-u^2}} \int_{\chi_{m}}^{\chi_{p}}
\frac{d\chi}{\sqrt{(\chi_{p}-\chi)(\chi-\chi_{m})(\chi-\chi_{n})}},
\\ \label{CQS} &&\mathcal{J}_1\equiv\frac{J_1}{\frac{\sqrt{\lambda}}{2\pi}}=\frac{1}{\sqrt{1-u^2}} \int_{\chi_{m}}^{\chi_{p}}
\frac{\left[1-v\left(v W-u
K\right)-\chi\right]d\chi}{\sqrt{(\chi_{p}-\chi)(\chi-\chi_{m})(\chi-\chi_{n})}},
\\ \nn &&\mathcal{J}_2\equiv\frac{J_1}{\frac{\sqrt{\lambda}}{2\pi}}=\frac{1}{\sqrt{1-u^2}} \int_{\chi_{m}}^{\chi_{p}}
\frac{\left(u\chi-vK\right)d\chi}{\sqrt{(\chi_{p}-\chi)(\chi-\chi_{m})(\chi-\chi_{n})}}.\eea

Computing the angular differences \bea\label{dad}
p\equiv\Delta\phi_1=\phi_1(r)-\phi_1(-r),\h
\tilde{p}\equiv\Delta\phi_2=\phi_2(r)-\phi_2(-r) ,\eea one finds
\bea\label{dp}&&p=\int_{-r}^{r}d\xi
f'_1=\frac{1}{\sqrt{1-u^2}}\int_{\chi_{m}}^{\chi_{p}} \left(\frac{v
W-u
K}{1-\chi}-v\right)\frac{d\chi}{\sqrt{(\chi_{p}-\chi)(\chi-\chi_{m})(\chi-\chi_{n})}},
\eea \bea\label{dd} &&\tilde{p}=\int_{-r}^{r}d\xi
f'_2=\frac{1}{\sqrt{1-u^2}}\int_{\chi_{m}}^{\chi_{p}}
\left(\frac{K}{\chi}-uv\right)\frac{d\chi}{\sqrt{(\chi_{p}-\chi)(\chi-\chi_{m})(\chi-\chi_{n})}}
.\eea

By using the formulas for the elliptic integrals given in Appendix
A, we can rewrite (\ref{CQS}), (\ref{dp}) and (\ref{dd}) in their
final form \bea\nn &&\mathcal{E}
=\frac{2(1-v^2)\sqrt{W}}{\sqrt{1-u^2}\sqrt{\chi_{p}-\chi_{n}}}
\mathbf{K}(1-\epsilon),
\\ \nn &&\mathcal{J}_1=\frac{2}{\sqrt{1-u^2}}\left(\frac{1-v\left(v W-u
K\right)-\chi_n}{\sqrt{\chi_{p}-\chi_{n}}}\mathbf{K}(1-\epsilon)-\sqrt{\chi_{p}-\chi_{n}}
\mathbf{E}(1-\epsilon)\right),
\\ \label{CQSf} &&\mathcal{J}_2=\frac{2}{\sqrt{1-u^2}} \left(u \sqrt{\chi_{p}-\chi_{n}}
\mathbf{E}(1-\epsilon)-\frac{vK-u\chi_n}{\sqrt{\chi_{p}-\chi_{n}}}\mathbf{K}(1-\epsilon)\right),
\\ \nn &&p=\frac{2}{\sqrt{1-u^2}\sqrt{\chi_{p}-\chi_{n}}}\left(\frac{v W-u K}{1-\chi_p}
\Pi\left(-\frac{\chi_{p}-\chi_{m}}{1-\chi_{p}}\vert
1-\epsilon\right)-v \mathbf{K}(1-\epsilon)\right), \\ \nn
&&\tilde{p}=\frac{2}{\sqrt{1-u^2}\sqrt{\chi_{p}-\chi_{n}}}\left(\frac{K}{\chi_p}
\Pi\left(1-\frac{\chi_{m}}{\chi_{p}}\vert 1-\epsilon\right) -u v
\mathbf{K}(1-\epsilon)\right) .\eea

\setcounter{equation}{0}
\section{Finite-Size Dyonic String}

In order to find the leading finite-size correction to the
energy-charge relation for the string with two angular momenta
(dyonic string), we have to consider the limit $\epsilon\to 0$ in
(\ref{3eqs}) and (\ref{CQSf}). The behavior of the complete elliptic
integrals in this limit is given in Appendix A. For the parameters
in the solution, we will use the ansatz \bea\nn
&&\chi_p=\chi_{p0}+\left(\chi_{p1}+\chi_{p2}\log(\epsilon)\right)\epsilon,
\\ \nn &&\chi_m=\chi_{m1}\epsilon,
\\ \nn &&\chi_n=\chi_{n1}\epsilon,
\\
\label{Dpars} &&v=v_0+\left(v_1+v_2\log(\epsilon)\right)\epsilon,
\\
\nn &&u=u_0+\left(u_1+u_2\log(\epsilon)\right)\epsilon,
\\ \nn &&W=1+W_{1}\epsilon,
\\ \nn &&K=K_{1}\epsilon.\eea
Replacing (\ref{Dpars}) into (\ref{3eqs}), one finds four equations
for the coefficients in the expansions of $\chi_p$ and $\chi_m$.
They are solved by \bea\nn &&\chi_{p0}=1-\frac{v_0^2}{1-u_0^2}, \\
\nn &&\chi_{p1}=
-\frac{1}{(1-u_0^2)^2(1-v_0^2-u_0^2)}\left[v_0\left(-2K_1
u_0(1-u_0^2)^2 \right.\right. \\ \label{chi}
&&+2(1-v_0^2-u_0^2)\left.\left.\left(v_0 u_0
u_1+(1-u_0^2)v_1\right)+v_0(1-u_0^2)(1-v_0^2-2u_0^2)W_1\right)\right],
\\ \nn &&\chi_{p2}=
-\frac{2v_0\left(v_0 u_0 u_2+(1-u_0^2)v_2\right)}{(1-u_0^2)^2} \\
\nn &&\chi_{m1}= -\frac{2v_0 u_0
K_1+(1-v_0^2)W_1}{1-v_0^2-u_0^2}-\chi_{n1}.\eea

As a next step, we impose the condition $p$ to be finite. It leads
to the relation \bea \label{pex} p=\arcsin\left(\frac{2
v_0\sqrt{1-v_0^2-u_0^2}}{1-u_0^2}\right),\eea as well as to four
equations for the parameters involved.

From the requirement $\mathcal{J}_2$ to be finite, one derives the
equality \bea \label{j2ex} \mathcal{J}_2=\frac{2
u_0\sqrt{1-v_0^2-u_0^2}}{1-u_0^2},\eea and two more equations.

The equations (\ref{pex}), (\ref{j2ex}) are solved by
\bea\label{zms}
v_0=\frac{\sin(p)}{\sqrt{\mathcal{J}_2^2+4\sin^2(p/2)}},\h
u_0=\frac{\mathcal{J}_2}{\sqrt{\mathcal{J}_2^2+4\sin^2(p/2)}}.\eea
After the replacement of (\ref{chi}) into the remaining six
equations, one finds that one of them is satisfied identically,
while the others can be solved with respect to $v_1$, $v_2$, $u_1$,
$u_2$, $W_1$, leading to the following form of the energy-charge
relation in the considered approximation \bea\label{echr1}
\mathcal{E}-\mathcal{J}_1=\frac{2\sqrt{1-v_0^2-u_0^2}}{1-u_0^2}
-\left(\frac{(1-v_0^2-u_0^2)^{3/2}}{2(1-u_0^2)} +
\sqrt{1-v_0^2-u_0^2}\chi_{n1}\right)\epsilon .\eea To the leading
order, the expansion for $\mathcal{J}_1$ gives \bea\label{eps}
\epsilon=16\exp\left(-\frac{2-\frac{2
v_0^2}{1-u_0^2}+\mathcal{J}_1\sqrt{1-v_0^2-u_0^2}}{1-v_0^2}\right).\eea
By using (\ref{zms}) and (\ref{eps}), (\ref{echr1}) can be rewritten
as \bea\label{IEJ1} &&\mathcal{E}-\mathcal{J}_1 =
\sqrt{\mathcal{J}_2^2+4\sin^2(p/2)} - \frac{16
\sin^2(p/2)\left(\sin^2(p/2)+2 \chi_{n1}\right)} {\sqrt{\mathcal{J}_2^2+4\sin^2(p/2)}}\\
\nn &&\exp\left[-\frac{2\left(\mathcal{J}_1 +
\sqrt{\mathcal{J}_2^2+4\sin^2(p/2)}\right)
\sqrt{\mathcal{J}_2^2+4\sin^2(p/2)}\sin^2(p/2)}{\mathcal{J}_2^2+4\sin^4(p/2)}
\right].\eea Thus, the new parameter in the dispersion relation is
$\chi_{n1}$ ($\chi_{n}=\chi_{n1}\epsilon$), reflecting the fact that
we are considering more general string solution. For
$\mathcal{J}_2=0$, (\ref{IEJ1}) simplifies to \bea\label{IEJ10}
\mathcal{E}-\mathcal{J}_1 =
2\sin\frac{p}{2}\left[1-4\left(\sin^2\frac{p}{2}+2
\chi_{n1}\right)e^{-2-\mathcal{J}_1\csc\frac{p}{2}}\right].\eea If
we set $\chi_{n1}=0$, (\ref{IEJ10}) reduces to the finite-size giant
magnon dispersion relation first found in \cite{AFZ06}, while
(\ref{IEJ1}) reproduces the result of \cite{HS08}.

Now, we are going to show how the equalities (\ref{IEJ10}) and
(\ref{IEJ1}) can be related to the ones found for giant magnons in
$TsT$-transformed $AdS_5\times S^5$ in \cite{BF08} and \cite{AB10}
respectively. To this end, let us consider the leading term in the
expansion for the angular difference $\tilde{p}$ \bea\label{pt1}
&&\tilde{p}=
-\frac{2K_1}{\chi_{n1}}\sqrt{\frac{1}{1-v_0^2-u_0^2+(1-u_0^2)\chi_{n1}}-\frac{1}{1-v_0^2-u_0^2}}
\\ \nn &&\times
\arctan\sqrt{-\frac{(1-u_0^2)\chi_{n1}}{1-v_0^2-u_0^2+(1-u_0^2)\chi_{n1}}}
+\frac{u_0
v_0}{\sqrt{1-v_0^2-u_0^2}}\log\left(\frac{\varepsilon}{16}\right).\eea
If instead of $\chi_{n1}$ we introduce the angle $\Phi$ as
\bea\label{ltd}
\frac{\Phi}{2}=\arctan\sqrt{-\frac{(1-u_0^2)\chi_{n1}}{1-v_0^2-u_0^2+(1-u_0^2)\chi_{n1}}},\eea
this gives \bea\label{kn1f}
\chi_{n1}=-\frac{1-v_0^2-u_0^2}{1-u_0^2}\sin^2(\Phi/2)
=-\sin^2(p/2)\sin^2(\Phi/2) ,\eea and the first term in (\ref{pt1})
takes the form \bea\label{K1}
\frac{2K_1(1-u_0^2)}{\left(1-v_0^2-u_0^2\right)^{3/2}}
\Phi\csc\Phi.\eea If we impose the natural condition (\ref{K1}) to
be an angle proportional to the angle $\Phi$, this gives
\bea\label{K12s}
K_{1}=\tilde{\Lambda}\frac{(1-v_0^2-u_0^2)^{3/2}}{2(1-u_0^2)}\sin\Phi
=\tilde{\Lambda}\frac{\sin^{4}(p/2)}{\sqrt{\mathcal{J}_2^2+4\sin^2(p/2)}}\sin\Phi,\eea
where $\tilde{\Lambda}$ does not depend on $\Phi$. In the
approximation under consideration, we do not see any criterion to
fix the parameter $\tilde{\Lambda}$. However in the next section, we
will show that $\tilde{\Lambda}=\pm 1$ must be fulfilled. As a
result, the relation between the angles $\tilde{p}$ and $\Phi$
becomes  \bea\label{fd} \tilde{p}+\mathcal{J}_2\frac{\mathcal{J}_1 +
\sqrt{\mathcal{J}_2^2+4\sin^2(p/2)}}{\mathcal{J}_2^2
+4\sin^4(p/2)}\sin(p)=\pm\Phi .\eea

In accordance with (\ref{kn1f}), the dispersion relation
(\ref{IEJ1}) takes the form \bea\label{IEJ2}
&&\mathcal{E}-\mathcal{J}_1 = \sqrt{\mathcal{J}_2^2+4\sin^2(p/2)} -
\frac{16 \sin^4(p/2)}
{\sqrt{\mathcal{J}_2^2+4\sin^2(p/2)}}\cos\Phi\\ \nn
&&\exp\left[-\frac{2\left(\mathcal{J}_1 +
\sqrt{\mathcal{J}_2^2+4\sin^2(p/2)}\right)
\sqrt{\mathcal{J}_2^2+4\sin^2(p/2)}\sin^2(p/2)}{\mathcal{J}_2^2+4\sin^4(p/2)}
\right],\eea which is the same as in \cite{AB10}. For
$\mathcal{J}_2=0$, (\ref{IEJ2}) reduces to the form found in
\cite{BF08}.

\setcounter{equation}{0}
\section{Subleading Correction}
Here our aim is to obtain the subleading correction to the
dispersion relation of a string with one angular momentum
$\mathcal{J}_1$, i.e. to (\ref{IEJ10}). Now, we will use the
following expansions for the parameters \bea\nn
&&\chi_p=\chi_{p0}+\left(\chi_{p1}+\chi_{p2}\log(\epsilon)\right)\epsilon
+
\left(\chi_{p20}+\chi_{p21}\log(\epsilon)+\chi_{p22}\log^2(\epsilon)\right)\epsilon^2,
\\ \nn &&\chi_m=\chi_{m1}\epsilon
+ \left(\chi_{m20}+\chi_{m21}\log(\epsilon)\right)\epsilon^2,
\\ \nn &&\chi_n=\chi_{n1}\epsilon
+ \left(\chi_{n20}+\chi_{n21}\log(\epsilon)\right)\epsilon^2,
\\
\label{slp} &&v=v_0+\left(v_1+v_2\log(\epsilon)\right)\epsilon
+\left(v_{20}+v_{21}\log(\epsilon)+v_{22}\log^2(\epsilon)\right)\epsilon^2,
\\
\nn &&u=\left(u_1+u_2\log(\epsilon)\right)\epsilon
+\left(u_{20}+u_{21}\log(\epsilon)+u_{22}\log^2(\epsilon)\right)\epsilon^2,
\\ \nn &&W=1+W_{1}\epsilon
+ \left(W_{20}+W_{21}\log(\epsilon)\right)\epsilon^2,
\\ \nn &&K=K_{1}\epsilon
+ \left(K_{20}+K_{21}\log(\epsilon)\right)\epsilon^2,\eea where
$u_0$ is set to zero in accordance with (\ref{zms}).

Replacing (\ref{slp}) into (\ref{3eqs}), one finds that one of the
equations leads to\footnote{Further on, we will use the plus sign.
We point out that the dispersion relation does not depend on this
sign.} \bea\nn K_1=\pm \sqrt{-\chi_{p0}\chi_{m1}\chi_{n1}}.\eea The
solutions of the remaining nine equations are \bea\nn
&&\chi_{p0}=1-v_0^2, \h\chi_{p1}=-v_0\left(2v_1+v_0W_1\right),
\h\chi_{p2}= -2v_0v_2, \h\chi_{m1}= -W_1-\chi_{n1},
\\ \nn &&\chi_{p20}= -\frac{1}{1-v_0^2}\left[v_1^2+v_0\left(2\left(1-v_0^2\right)\left(v_1W_1+v_{20}\right)
-v_0^3\left(u_1^2+W_{20}\right)\right.\right.
\\ \nn
&&-\left.\left.
2u_1\sqrt{\left(1-v_0^2\right)\chi_{n1}\left(W_1+\chi_{n1}\right)}+
v_0\left(u_1^2-v_1^2+W_{20}+\chi_{n1}\left(W_1+\chi_{n1}\right)\right)\right)\right],
\\ \nn &&\chi_{p21}= \frac{1}{1-v_0^2}\left[-\left(1-v_0^2\right) \left(2v_0v_2W_1
+2v_1v_2+v_0\left(2v_0u_1u_2+2v_{21}+v_0W_{21}\right),\right)\right.
\\ \nn
&&+\left.2v_0u_2\sqrt{\left(1-v_0^2\right)\chi_{n1}\left(W_1+\chi_{n1}\right)}\right],
\\ \nn &&\chi_{p22}=-\left(v_0^2u_2^2+v_2^2+2v_0v_{22}\right),
\\ \nn &&\chi_{m20}=-\frac{1}{1-v_0^2}\left[\left(1-v_0^2\right)\left(W_{20}+\chi_{n20}\right)
-v_0^2\chi_{n1}\left(W_1+\chi_{n1}\right)\right.
\\ \nn &&\left.+2v_0u_1\sqrt{\left(1-v_0^2\right)\chi_{n1}\left(W_1+\chi_{n1}\right)}\right],
\\ \nn &&\chi_{m21}=-\frac{1}{1-v_0^2}\left[\left(1-v_0^2\right)\left(W_{21}+\chi_{n21}\right)
+2v_0u_2\sqrt{\left(1-v_0^2\right)\chi_{n1}\left(W_1+\chi_{n1}\right)}\right],\eea
where \bea\nn
K_1=\sqrt{\left(1-v_0^2\right)\chi_{n1}\left(W_1+\chi_{n1}\right)}.\eea

From the condition $p$ to be finite, we obtain \bea \nn
p=\arcsin\left(2 v_0\sqrt{1-v_0^2}\right),\eea solved by \bea\nn
v_0=\cos\frac{p}{2},\eea and nine more equations as well. Three of
them can be solved independently, leading to \bea\nn
&&v_1=\frac{1}{4}v_0\left(1-v_0^2+2\chi_{n1}\right)\left(1-\log(16)\right),
\\ \nn &&v_2=\frac{1}{4}v_0\left(1-v_0^2+2\chi_{n1}\right),
\\ \nn &&W_1=-\left(1-v_0^2+2\chi_{n1}\right).\eea

Our next condition $\mathcal{J}_2=0$, gives \bea\nn
u_1=\frac{v_0K_1}{1-v_0^2}\log(4),\h
u_2=-\frac{v_0K_1}{2(1-v_0^2)},\eea where \bea\nn
K_1=\sqrt{-\left(1-v_0^2\right)\left(1-v_0^2+\chi_{n1}\right)\chi_{n1}},\eea
and another three equations. Thus, a system of nine equations
remains to be solved, using the results obtained so far. It turns
out that one of them is satisfied identically, while the others can
be solved with respect to $v_{20}$, $v_{21}$, $v_{22}$, $u_{20}$,
$u_{21}$, $u_{22}$, $W_{20}$, $W_{21}$.

The parameters $\chi_{n20}$, $\chi_{n21}$, $K_{20}$, $K_{21}$, are
still undetermined. To fix them, let us consider the expansion for
the angular difference $\tilde{p}$. To the leading order, we have
\bea\nn
\tilde{p}=2\arctan\sqrt{\frac{-\chi_{n1}}{1-v_0^2+\chi_{n1}}}, \eea
solved by \bea\nn \chi_{n1}=-(1-v_0^2)\sin^2\frac{\tilde{p}}{2}
=-\sin^2\frac{p}{2}\sin^2\frac{\tilde{p}}{2},\eea and leading to
\bea\label{k1} K_1=\frac{1}{2}\sin^3\frac{p}{2}\sin\tilde{p}.\eea
Comparing (\ref{k1}) with (\ref{K12s}) taken at $\mathcal{J}_2=0$,
one finds $\tilde{\Lambda}=1$, $\tilde{p}=\Phi$ (see also
(\ref{fd})).

Now, we impose the condition the equality $\tilde{p}=\Phi$ to be
valid in the subleading order also. This gives four equations, which
can be solved with respect to $\chi_{n20}$, $\chi_{n21}$, $K_{20}$
and $K_{21}$. The solutions for all parameters in their final form
are given in appendix A.

Replacing all solutions for the parameters appearing in the
expansion for $\mathcal{E}-\mathcal{J}_1$, we find the following
dispersion relation up to the subleading order correction
\bea\label{IEJ10SL} &&\mathcal{E}-\mathcal{J}_1 =
2\sin\frac{p}{2}\left(1-4\cos\tilde{p}\
\sin^2\frac{p}{2}e^{-2-\mathcal{J}_1\csc\frac{p}{2}}\right)
\\ \nn
&&-\left\{4\sin^3\frac{p}{2}\left[13+9\cos p
+16\cos\tilde{p}+(1+3\cos p)\cos 2\tilde{p}\right.\right.
\\ \nn
&&-64\left.\left.(1-\log2)\log 2 (1+\cos
p)\sin^2\tilde{p}\right]\right.
\\ \nn
&&+4\mathcal{J}_1\left.\left(3-2\cos p-\log 16+(\log 16
-1)\left(\cos 2\tilde{p}+2\cos 2p \
\sin^2\tilde{p}\right)\right)\right.
\\ \nn
&&- 16\mathcal{J}_1^2  \sin\frac{p}{2}\ \cos^2\frac{p}{2} \ \cos
2\tilde{p}\left.\right\}e^{-4-2\mathcal{J}_1\csc\frac{p}{2}},\eea
where $\tilde{p}\ne 0$, describes the deviation of the energy-charge
relation corrections from the ordinary giant magnon case. This is
our final result for a finite-size string with one (large) angular
momentum.

\setcounter{equation}{0}
\section{Concluding Remarks}
In this paper we considered the most general string configurations
on the $R_t\times S^3$ subspace of $AdS_5\times S^5$, described by
the NR integrable system. Imposing appropriate conditions on the
parameters involved, we restrict ourselves to string solutions,
which contain as particular cases the finite-size giant magnons with
one and two angular momenta. Taking the limit in which the modulus
of the elliptic integrals is close to one, we found the corrections
to the dispersion relations. For the dyonic string, this is done to
the leading order. For the string with one nonzero angular momentum,
the subleading correction is also obtained. In both cases, the
corrections depend on one more parameter $\Phi$ or $\tilde{p}$. When
$\Phi=0$, the finite-size giant magnons energy-charge relations
should be reproduced. It is true to the leading order. However, this
is not the case, when we consider the subleading correction. Indeed,
according to \cite{AFZ06}, in conformal gauge, it is given by
\bea\label{AFZr} -8\sin\frac{p}{2}\left[\left(7+6\cos
p\right)\sin^2\frac{p}{2} +2\mathcal{J}_1\left(2+3\cos
p\right)\sin\frac{p}{2}
+2\mathcal{J}_1^2\cos^2\frac{p}{2}\right]e^{-4-2\mathcal{J}_1\csc\frac{p}{2}}
,\eea while from (\ref{IEJ10SL}) we obtain \bea\label{OURr}
-8\sin\frac{p}{2}\left[3\left(5+2\cos p\right)\sin^2\frac{p}{2}
+2\mathcal{J}_1\sin\frac{p}{2}
-2\mathcal{J}_1^2\cos^2\frac{p}{2}\right]e^{-4-2\mathcal{J}_1\csc\frac{p}{2}}
.\eea Thus, only the terms containing $\mathcal{J}_1^2$ coincide up
to a sign. Similar discrepancy was observed in \cite{KMcL08} too,
where by using different approach, the authors found agreement with
(\ref{AFZr}) for the leading term $\propto \mathcal{J}_1^2$, but
disagreement at the subleading in $\mathcal{J}_1$ orders. In
(\ref{OURr}), we have additional sign difference. At the moment, we
do not know the reason, and can not explain why these three results
are different. Obviously, this issue should be studied in more
detail.

The approach we used here, can be applied to find the finite-size
effects on the dispersion relations for different string
configurations in $AdS_5\times S^5$ (generalized spiky strings for
example), as well as the finite-size effects in other backgrounds.
In particular, this problem for dyonic giant magnons in the
$\gamma$-deformed $AdS_4\times CP^3_{\gamma}$ is under investigation
\cite{AB11}.

\section*{Acknowledgements}
This work was supported in part by NSFB VU-F-201/06 and DO 02-257
grants.

\def\theequation{A.\arabic{equation}}
\setcounter{equation}{0}
\begin{appendix}

\section{Elliptic Integrals, $\epsilon$-Expansions and Solutions for the Parameters}

The elliptic integrals appearing in the main text are given by
\bea\nn &&\int_{\chi_{m}}^{\chi_{p}}
\frac{d\chi}{\sqrt{(\chi_{p}-\chi)(\chi-\chi_{m})(\chi-\chi_{n})}}
=\frac{2}{\sqrt{\chi_{p}-\chi_{n}}}\mathbf{K}(1-\epsilon),\\ \nn
&&\int_{\chi_{m}}^{\chi_{p}} \frac{\chi
d\chi}{\sqrt{(\chi_{p}-\chi)(\chi-\chi_{m})(\chi-\chi_{n})}}
\\ \nn &&
=\frac{2\chi_{n}}{\sqrt{\chi_{p}-\chi_{n}}}\mathbf{K}(1-\epsilon)+2\sqrt{\chi_{p}-\chi_{n}}
\mathbf{E}(1-\epsilon),
\\ \nn &&\int_{\chi_{m}}^{\chi_{p}}
\frac{d\chi}{\chi\sqrt{(\chi_{p}-\chi)(\chi-\chi_{m})(\chi-\chi_{n})}}
=\frac{2}{\chi_{p}\sqrt{\chi_{p}-\chi_{n}}}
\Pi\left(1-\frac{\chi_{m}}{\chi_{p}}\vert 1-\epsilon\right),
\\ \nn&&\int_{\chi_{m}}^{\chi_{p}}
\frac{d\chi}{\left(1-\chi\right)\sqrt{(\chi_{p}-\chi)(\chi-\chi_{m})(\chi-\chi_{n})}}
\\ \nn &&=\frac{2}{\left(1-\chi_{p}\right)\sqrt{\chi_{p}-\chi_{n}}}
\Pi\left(-\frac{\chi_{p}-\chi_{m}}{1-\chi_{p}}\vert
1-\epsilon\right),\eea where \bea\nn
\epsilon=\frac{\chi_{m}-\chi_{n}}{\chi_{p}-\chi_{n}}.\eea

We use the following expansions for the complete elliptic integrals
\cite{w} \bea\nn &&\mathbf{K}(1-\epsilon)=
-\frac{1}{2}\log\left(\frac{\epsilon}{16}\right)-\frac{1}{4}\left(1+\frac{1}{2}\log\left(\frac{\epsilon}{16}\right)\right)
\epsilon -\frac{3}{128}\left(7+3
\log\left(\frac{\epsilon}{16}\right)\right)\epsilon^2 +\ldots,
\\ \nn
&&\mathbf{E}(1-\epsilon)=
1-\frac{1}{4}\left(1+\log\left(\frac{\epsilon}{16}\right)\right)\epsilon-\frac{1}{64}\left(13+6
\log\left(\frac{\epsilon}{16}\right)\right)\epsilon^2 +\ldots,
\\ \nn &&\Pi(-n|1-\epsilon)=\frac{2\sqrt{n}\arctan(\sqrt{n})-\log\left(\frac{\epsilon}{16}\right)}{2(1+n)}
-\frac{2-4\sqrt{n}\arctan(\sqrt{n})+(1-n)\log\left(\frac{\epsilon}{16}\right)}{8(1+n)^2}\epsilon
\\ \nn &&-\frac{21-12 n-5 n^2-48\sqrt{n}\arctan(\sqrt{n})
+3\left(3-(6+n)n\right)\log\left(\frac{\epsilon}{16}\right)}{128(1+n)^3}\epsilon^2
+\ldots, \h n>0.\eea

We use also the equality \cite{PBM-III} \bea\nn
\Pi(\nu|m)=\frac{q_{1}}{q}\Pi(\nu_1|m)-\frac{m}{q\sqrt{-\nu\nu_1}}\mathbf{K}(m),\eea
where \bea\nn &&q=\sqrt{(1-\nu)\left(1-\frac{m}{\nu}\right)},\h
q_1=\sqrt{(1-\nu_1)\left(1-\frac{m}{\nu_1}\right)},\\ \nn
&&\nu=\frac{\nu_1-m}{\nu_1-1},\h \nu_1<0,\h m<\nu<1 .\eea

For the parameters, we use the following expansions  \bea\nn
&&\chi_p=\chi_{p0}+\left(\chi_{p1}+\chi_{p2}\log(\epsilon)\right)\epsilon
+
\left(\chi_{p20}+\chi_{p21}\log(\epsilon)+\chi_{p22}\log^2(\epsilon)\right)\epsilon^2,
\\ \nn &&\chi_m=\chi_{m1}\epsilon
+ \left(\chi_{m20}+\chi_{m21}\log(\epsilon)\right)\epsilon^2,
\\ \nn &&\chi_n=\chi_{n1}\epsilon
+ \left(\chi_{n20}+\chi_{n21}\log(\epsilon)\right)\epsilon^2,
\\
\nn &&v=v_0+\left(v_1+v_2\log(\epsilon)\right)\epsilon
+\left(v_{20}+v_{21}\log(\epsilon)+v_{22}\log^2(\epsilon)\right)\epsilon^2,
\\
\nn &&u=u_0+\left(u_1+u_2\log(\epsilon)\right)\epsilon
+\left(u_{20}+u_{21}\log(\epsilon)+u_{22}\log^2(\epsilon)\right)\epsilon^2,
\\ \nn &&W=1+W_{1}\epsilon
+ \left(W_{20}+W_{21}\log(\epsilon)\right)\epsilon^2,
\\ \nn &&K=K_{1}\epsilon
+ \left(K_{20}+K_{21}\log(\epsilon)\right)\epsilon^2.\eea

Considering the case of a string with two angular momenta, we
neglect the terms proportional to $\epsilon^2$, thus obtaining the
leading correction to the dispersion relation only. To find the
subleading correction for a string with one angular momentum, we set
$u_0=0$.

The solutions for the parameters for the first case,
$\mathcal{J}_2\ne 0$, are given by ($\tilde{\Lambda}=1$) \bea\nn
&&\chi_{p0}= \sin^2(p/2),
\\ \nn &&\chi_{p1}=
-\frac{1}{16\left(\mathcal{J}_2^2+4\sin^4(p/2)\right)}
\left[\frac{1}{\mathcal{J}_2^2+4\sin^2(p/2)} \left(\cos\Phi
\left(\cos p \left(15+8\mathcal{J}_2^2+60\log
2\right)\right.\right.\right.
\\ \nn
&&-\left.\left.\left. (1+\log
16)\left(10+\left(6+\mathcal{J}_2^2\right)\cos 2p -
\cos3p\right)\right.\right.\right.
\\ \nn
&&+\left.\left.\left.\mathcal{J}_2^2\left(\log 16
-7+\mathcal{J}_2^2\left(\log 256
-2\right)\right)\right)\sin^2p\right)\right.
\\ \nn
&&-\left.\frac{16\mathcal{J}_2\sin\Phi \sin^3(p/2)\cos
(p/2)\left(5+2\mathcal{J}_2^2-4\cos p -\cos 2p\right)\log
2}{\sqrt{\mathcal{J}_2^2+4\sin^2(p/2)}}\right],
\\ \nn &&\chi_{p2}=
\frac{1}{8\left(\mathcal{J}_2^2+4\sin^2(p/2)\right)
\left(\mathcal{J}_2^2+4\sin^4(p/2)\right)} \left[\sin(p/2)\sin p
\left(\cos(p/2)\left(\mathcal{J}_2^2 \right.\right.\right.
\\ \nn &&+2 \mathcal{J}_2^4 -10+15 \left.\left.\left.\cos p -\left(6+\mathcal{J}_2^2\right)\cos 2p +\cos
3p\right)\cos\Phi \right.\right.
\\ \nn
&&-\left.\left.\mathcal{J}_2\sqrt{\mathcal{J}_2^2+4\sin^2(p/2)}
\left(5+2\mathcal{J}_2^2 - 4\cos p -\cos 2p\right)\sin(p/2)\sin
\Phi\right)\right],
\\ \nn &&\chi_{m1}= \sin^2(p/2)\cos^2(\Phi/2),
\\ \nn &&\chi_{n1}= -\sin^2(p/2)\sin^2(\Phi/2), \eea

\bea\nn &&v_0= \frac{\sin p}{\sqrt{\mathcal{J}_2^2+4\sin^2(p/2)}},
\\ \nn &&v_1= \frac{\sin^2(p/2)}{16\left(\mathcal{J}_2^2+4\sin^2(p/2)\right)^{3/2}
\left(\mathcal{J}_2^2+4\sin^4(p/2)\right)}
\\\nn &&\times
\left[\cos\Phi \left(2\left(7-28\log 2 +4\mathcal{J}_2^2\left(\log
128 -3+2\mathcal{J}_2^2\left(\log 4 -1\right)\right)\right)\sin p
\right.\right.
\\ \nn
&&+\left.\left. 2\left(28\log 2 -7-2\mathcal{J}_2^2\left(\log 16
-3\right)\right)\sin 2p -2\left(\log 4096 -3+\mathcal{J}_2^2 \log
16\right)\sin 3p \right.\right.
\\ \nn
&&+\left.\left. \left(\log 16 -1\right)\sin 4p \right)
+2\mathcal{J}_2\sqrt{\mathcal{J}_2^2+4\sin^2(p/2)} \left(14-32\log 2
+4\mathcal{J}_2^2+2\cos 2p \right.\right.
\\ \nn
&&+\cos p \left.\left.\left(36\log 2 -17+4\mathcal{J}_2^2\left(\log
16 -1\right)\right) -\cos 3p \left(\log 16 -1\right)\right)
\sin\Phi\right],
\\ \nn &&v_2= -\frac{\sin^2(p/2)}{8\left(\mathcal{J}_2^2+4\sin^2(p/2)\right)^{3/2}
\left(\mathcal{J}_2^2+4\sin^4(p/2)\right)}
\left[\left(-10+6\mathcal{J}_2^2+4\mathcal{J}_2^4\right.\right.
\\ \nn &&+\left.\left. \left(15-4\mathcal{J}_2^2\right)\cos p
-2\left(3+\mathcal{J}_2^2\right)\cos 2p + \cos 3p\right)\sin p \
\cos \Phi\right.
\\ \nn
&&-\left.\mathcal{J}_2\sqrt{\mathcal{J}_2^2+4\sin^2(p/2)}
\left(8-\left(9+4\mathcal{J}_2^2\right)\cos p +\cos 3p\right)\sin
\Phi\right], \eea

\bea\nn &&u_0=
\frac{\mathcal{J}_2}{\sqrt{\mathcal{J}_2^2+4\sin^2(p/2)}},
\\ \nn &&u_1= \frac{\sin^3(p/2)}{2\left(\mathcal{J}_2^2+4\sin^2(p/2)\right)^{3/2}
\left(\mathcal{J}_2^2+4\sin^4(p/2)\right)}
\left[\mathcal{J}_2\cos\Phi\left(3+2\mathcal{J}_2^2-4\cos p
\right.\right.
\\ \nn &&+\left.\left.\cos 2p +4\left(-5-2\mathcal{J}_2^2+4\cos p +\cos 2p\right)
 \log 2\right)\sin(p/2)\right.
\\ \nn &&+\left. 4\cos(p/2)\sqrt{\mathcal{J}_2^2+4\sin^2(p/2)}
\left(3-2\mathcal{J}_2^2-4\cos p +\cos 2p\right)\log 2 \
\sin\Phi\right],
\\ \nn &&u_2= \frac{\sin^2(p/2)}{2\left(\mathcal{J}_2^2+4\sin^2(p/2)\right)^{3/2}
\left(\mathcal{J}_2^2+4\sin^4(p/2)\right)}
\left[\mathcal{J}_2\left(5+2\mathcal{J}_2^2-4\cos p -\cos
2p\right)\right.
\\ \nn && \times\left. \sin^2(p/2)\cos\Phi -\frac{1}{2}\sqrt{\mathcal{J}_2^2+4\sin^2(p/2)}
\left(3-2\mathcal{J}_2^2-4\cos p +\cos 2p\right)\sin p \
\sin\Phi\right], \eea

\bea\nn &&W_1= -4 \sin^5(p/2)\frac{\sin(p/2)\cos\Phi+
\mathcal{J}_2\frac{\cos(p/2)}{\sqrt{\mathcal{J}_2^2+4\sin^2(p/2)}}\sin\Phi}{\mathcal{J}_2^2+4\sin^4(p/2)},
\\ \nn &&K_1= \frac{\sin^4(p/2)}{\sqrt{\mathcal{J}_2^2+4\sin^2(p/2)}}\sin\Phi. \eea
In the equalities above, the angle $\Phi$ is related to the angle
$\tilde{p}$, defined in (\ref{dad}), according to \bea\nn \Phi=
\tilde{p}+\mathcal{J}_2\frac{\mathcal{J}_1 +
\sqrt{\mathcal{J}_2^2+4\sin^2(p/2)}}{\mathcal{J}_2^2
+4\sin^4(p/2)}\sin p .\eea

The solutions for the parameters for the second case,
$\mathcal{J}_2= 0$, are as follows \bea\nn &&\chi_{p0}=
\sin^2\frac{p}{2}, \h \chi_{p1}= \frac{1}{8}\cos\tilde{p}\ \sin^2 p
\left(1+\log
16\right), \h \chi_{p2}= -\frac{1}{8}\cos\tilde{p}\ \sin^2 p , \\
\nn &&\chi_{p20}= \frac{1}{512}\left[\cos 3p \left(-1+8\log 2
\left(-1+\log 4 \left(-3+\log 16\right)\right)\right)\right.
\\ \nn &&+32\left.\log 2 \left(-1-4\log^2 2 +\cos 2p\left(-1+\log 4\right)^2
+\log 16 \right.\right. \\ \nn &&
+2\left.\left.\left(2\cos\tilde{p}+\cos 2\tilde{p}\left(-1+\log
4\right)^2\right)\sin^2p\right)+\cos p \left(1+16\left(3-\log
16\right)\log^22+\log 256 \right.\right. \\ \nn &&+8\left.\left.\cos
2\tilde{p} \left(1+8\log^22\left(3+\log 16\right)+\log
256\right)\sin^2p\right)\right],
\\ \nn &&\chi_{p21}=
\frac{1}{128}\left[2\cos^2\frac{p}{2}\left(5-8\cos\tilde{p}+\cos 2p
\left(1-12\log 2\left(-1+\log 4\right)\right)\right.\right.
\\ \nn
&&+2\cos p \left.\left.\left(-3+\log 16 +4\cos\tilde{p}\right)+\log
16\left(-5+\log 64\right)\right)\right.
\\ \nn
&&-4\left.\cos 2\tilde{p}\left(1+\log 16\left(-2+\log8\right)+\cos p
\left(1+12\log^22+\log 64\right)\right)\sin^2p\right],
\\ \nn &&\chi_{p22}=\frac{1}{128}\sin^2p\left[\cos p
\left(4-8\log 2+\cos 2\tilde{p}\left(2+\log 256\right)\right)
-2\left(-5+\log 256\right)\sin^2\tilde{p}\right], \eea

\bea\nn &&\chi_{m1}= \cos^2\frac{\tilde{p}}{2}\sin^2\frac{p}{2}, \\
\nn &&\chi_{m20}= \frac{1}{64}\left[3+\cos 2p
\left(-1+\log16\left(-1+\log 4\right)\right)+\log 4096 \right.
\\ \nn
&&+ 8\left.\log^22\left(-1+4\cos 2\tilde{p} \
\sin^2\frac{p}{2}\right) +4\cos\tilde{p}\left(1+\log 16\right)\sin^2
p\right.
\\ \nn
&&+2\left.\cos p \left(-1-\log16 +2\cos 2\tilde{p}\left(1+8\log^22 +
\log 16\right)\sin^2\frac{p}{2}\right)\right],
\\ \nn
&&\chi_{m21}= -\frac{1}{4}
\cos^2\frac{\tilde{p}}{2}\sin^2\frac{p}{2}\left[1+\cos p \
\cos\tilde{p}-\left(1+\cos p\right)\left(1-\cos \tilde{p}\right)\log
16\right],\eea

\bea \nn &&\chi_{n1}=
-\sin^2\frac{\tilde{p}}{2}\sin^2\frac{p}{2},
\\ \nn &&\chi_{n20}=-\frac{1}{2}\sin^2\frac{\tilde{p}}{2}
\left[1+\log 16 \left(\log 4 +\cos p \left(1+\log 4
\right)\right)\right.
\\ \nn &&-\left.\cos^4\frac{p}{2}\left(1+16\log^22+\log 16\right)
-\left(8\log^22+\cos
p\left(1+8\log^22+\log16\right)\right)\sin^2\frac{p}{2}\sin^2\frac{\tilde{p}}{2}\right]
\\ \nn &&\chi_{n21}=- \frac{1}{4}\left[1-\cos p \ \cos\tilde{p} -\left(1+\cos p\right)
\left(1+\cos \tilde{p}\right)\log
16\right]\sin^2\frac{p}{2}\sin^2\frac{\tilde{p}}{2},\eea

\bea\nn
&&v_0=\cos\frac{p}{2},\h v_1=\frac{1}{4} \left(1-\log 16\right)\cos
\tilde{p} \ \cos\frac{p}{2} \sin^2\frac{p}{2},\h v_2=\frac{1}{4}\cos
\tilde{p} \ \cos\frac{p}{2} \ \sin^2\frac{p}{2},
\\ \nn &&v_{20}=\frac{1}{512}\cos\frac{p}{2}\left[-37+\cos2p \left(5-8\log 2
\left(7+\log 16 \left(-5+\log 16\right)\right)\right)\right.
\\ \nn &&+8\left.\log 2\left(23-36\log 2+\log^2 16\right)+4\left(\cos 2\tilde{p}
\left(19-8\log 2\left(13+8\log 2\left(-3+\log
4\right)\right)\right)\right.\right. \\ \nn
&&-32\left.\left.\cos\tilde{p}\left(-1+\log
4\right)\right)\sin^2\frac{p}{2}\right. \\ \nn &&+4\left.\cos
p\left(8\left(-1+\log 4\right)^2+\cos 2\tilde{p}\left(1-128\log^3 2
+\log 256\right)\sin^2\frac{p}{2}\right)\right], \\ \nn
&&v_{21}=\frac{1}{256}\cos\frac{p}{2} \left[-23-24\log
2\left(-3+\log 4\right)+\cos 2p\left(7+8\log 2\left(-5+\log
64\right)\right)\right. \\ \nn &&+\left. 4\left(8\cos\tilde{p}+\cos
2\tilde{p}\left(13+48\left(-1+\log 2\right)\log
2\right)\right)\sin^2\frac{p}{2}\right. \\ \nn &&+\left. 4\cos p
\left(4-4\log 4+\cos
2\tilde{p}\left(-1+48\log^22\right)\sin^2\frac{p}{2}\right)\right],
\\ \nn &&v_{22}=-\frac{1}{128}\cos\frac{p}{2} \left[12+\cos\left(p+2\tilde{p}\right)+
8\left(-2+\cos\left(p+2\tilde{p}\right)\right)\log 2-8\cos
p\left(-1+\log 4\right)\right.
\\ \nn &&+2\left.\cos 2\tilde{p}\left(-5+\log 256\right)
 + \cos\left(p-2\tilde{p}\right)\left(1+\log 256\right)\right]\sin^2\frac{p}{2},
\eea

\bea\nn &&u_1=\frac{1}{2}\log 2 \ \sin p \ \sin \tilde{p},\h
u_2=-\frac{1}{8}\sin p \ \sin \tilde{p}, \\ \nn
&&u_{20}=\frac{1}{16}\sin p\left[\left(-1+\log
16\right)\sin\tilde{p}\right.
\\ \nn &&+\log 4 \left. \left(\left(-1+\log 4\right)^2+2\cos p \log 2 (\left(1+\log
4\right)\right)\sin 2\tilde{p}\right], \\ \nn
&&u_{21}=\frac{1}{32}\sin p \left[-2\sin\tilde{p}-\left(1+\log 16
\left(-2+\log 8\right) +4\cos p \log 2 \left(1+\log
8\right)\right)\sin 2\tilde{p}\right], \\ \nn
&&u_{22}=\frac{1}{64}\left[-2+\cos p +4\left(1+\cos p\right)\log
2\right]\sin p \ \sin 2\tilde{p},\eea

\bea\nn &&W_1=-\cos\tilde{p} \
\sin^2\frac{p}{2}, \\ \nn
&&W_{20}=\frac{1}{64}\left[11-16\cos\tilde{p}+\log 16 \left(-7+\log
16\right)+\cos 2p \left(1-16\log^22 +\log 4096\right)\right.
\\ \nn &&-\left. 4\cos 2\tilde{p}\left(1+\log 16 \left(-1+\log 16\right)\right)
\sin^2\frac{p}{2}\right.
\\ \nn &&+\left. 4\cos p \left(-3+4\cos\tilde{p}+\log 16 -
\cos 2\tilde{p}\left(3+16\log^22+\log
16\right)\sin^2\frac{p}{2}\right)\right],
 \\ \nn
&&W_{21}= \frac{1}{16}\sin^2\frac{p}{2}\left[5+3\cos p - \cos
2\tilde{p}+\cos p \ \cos 2\tilde{p}-16\left(1+\cos p\right)\log 2 \
\sin^2\tilde{p}\right],\eea

\bea\nn
&&K_1=\frac{1}{2}\sin^3\frac{p}{2} \ \sin\tilde{p},
\\ \nn &&K_{20}= \frac{1}{8}\sin\frac{p}{2} \ \sin\tilde{p}\left[2+8\log^22+\log 16 +\cos p \left(1+8\log^22+\log 256\right)
\right. \\ \nn &&-\left. \cos^4\frac{p}{2}\left(3+\log^2 16+\log
4096\right)\right.
\\ \nn &&\left.
-\left(1+\log 16+\log^2 16 +\cos p \left(3+\log^2 16+\log
4096\right)\right)\sin^2\frac{p}{2}\sin^2\frac{\tilde{p}}{2}\right],
\\ \nn &&K_{21}= -\frac{1}{16}\sin^3\frac{p}{2} \ \sin\tilde{p} \ \cos\tilde{p}
\left[1+\log 256 +\cos p \left(3+\log 256\right)\right].\eea

\end{appendix}

\end{document}